\documentclass[12pt]{article}

\advance\hoffset by -1.1truecm
\advance\voffset by -1.0truecm

\def\be{\begin{equation}}
\def\ee{\end{equation}}
\def\ba{\begin{eqnarray}}
\def\ea{\end{eqnarray}}

\newcommand{\Z}{\:\mbox{\sf Z} \hspace{-0.82em} \mbox{\sf Z}\,}
\newcommand{\N}{\mbox{\rm N} \hspace{-0.9em} \mbox{\rm I}\,\,\,}
\newcommand{\R}{\mbox{I \hspace{-0.82em} R}}
\newcommand{\tphi}{{\tilde{\phi}}}

\newcommand{\B}{ {B_{\omega_{max}}} }
\newcommand{\oT}{ {\overline{T}(\phi)} }
\newcommand{\dT}{{{\Delta T(\phi)}}}

\newcommand{\Hi}{ {H_{\omega_{max}}} }
\newcommand{\ve}{ {{\bf t}} }

\newcommand{\sn}{\smallskip\newline}
\newcommand{\mn}{\medskip\newline}
\newcommand{\bn}{\bigskip\newline}

\begin{document}
\title{Black Holes, Bandwidths and Beethoven}
\author{Achim Kempf\\
Institute for Fundamental Theory, Department of Physics\\
University of Florida, Gainesville, FL 32611, USA\\
{\small Email:  kempf@phys.ufl.edu}}

\date{}

\maketitle

\vskip-7.5truecm

\hskip11.7truecm
{\tt UFIFT-HEP-99-09}

\hskip11.7truecm {\tt gr-qc/9907084} \vskip7.1truecm

\begin{abstract}
It is usually believed that  a
function $\phi(t)$ whose Fourier spectrum is
bounded can vary at most as fast as its highest frequency
 component $\omega_{max}$.
This is in fact not the case, as Aharonov, Berry and others
drastically demonstrated with explicit
counter examples, so-called superoscillations. 
It has been claimed
that even the recording of
an entire  Beethoven symphony can occur as part of a
signal with 1Hz bandwidth. 
Bandlimited functions also occur as ultraviolet regularized fields.
Their superoscillations have been suggested, for example, to resolve
the transplanckian frequencies problem
of black hole radiation.

Here, we give an exact proof for generic
superoscillations. Namely,
we show that for every fixed bandwidth there exist
 functions which pass
through any
finite number of arbitrarily prespecified points.
Further, we show that, in spite of the presence of
superoscillations, the behavior of bandlimited
functions can be characterized reliably, namely
through an uncertainty relation:
The standard deviation $\Delta T$ of samples
$\phi(t_n)$ taken at the
Nyquist rate obeys: $ \Delta T \ge 1/4\omega_{max}$.
This uncertainty relation generalizes to variable
bandwidths. We identify the bandwidth as the 
in general spatially variable
finite local density of the degrees of freedom
of ultraviolet regularized fields. 
\end{abstract}
$\qquad ~$ PACS: 11.10.-z, 02.30.Sa, 04.70.Dy, 89.70.+c

\newpage

\section{Introduction}
%%%%%%%%%%%%%%%%%%%%%%%%%%%%%%%%%%%%%%%%%%%%%%%%%
Functions which contain only frequencies up to a certain
maximum frequency
occur in various contexts from theoretical physics to
the applied sciences. For example, in quantum field theory 
the method of  ``ultraviolet" regularization 
by energy-momentum cut-off yields fields which are frequency limited.
Frequency limited functions also occur  for example 
as so-called ``bandlimited signals"
in communication engineering.
\sn
Intuitively, one may expect that a frequency limited function,
$\phi(t)$, can vary at most as fast as its highest frequency
component, $\omega_{max}$.
In fact, this is not the case.
\sn
Aharonov et al \cite{aharonovetal} and Berry
 \cite{berry},
gave explicit examples - which they named
superoscillations - which
drastically demonstrate that
frequency limited functions are able to oscillate for
 arbitrarily long finite
intervals arbitrarily faster than
the highest frequency component which they contain.
It has even been conjectured, in \cite{berry}, that
for example 5000 seconds of
a 20 KHz bandwidth recording of a symphony of
Beethoven can be part of a
1Hz  band\-limited signal.  
\bn
Before we begin our investigation of superoscillations a few remarks
on our use of terminology may be in place.
\sn
Frequency limited 
functions and superoscillations 
not only occur in
ultraviolet cut-off quantum field theory 
and in information theory but also
in a whole variety of other physical contexts.
Superoscillation are known to occur for example with
evanescent waves and quantum billiards,
see \cite{berry2}, and for example with effects of apparent superluminal
propagation in unstable media such as media with inverted level
populations, see e.g. \cite{aharezste}. 
We will here mostly be concerned with the
 general properties of superoscillations
and we could therefore use as our terminology  
the language of any one of these contexts where superoscillations occur. 
\sn
Our choice of terminology here will be to use both the language of 
quantum field theory and the language of information theory.
\sn
We make this choice because, on the one hand,  
our main interest is in the
implications of superoscillations in ultraviolet 
regular quantum field theories.
On the other hand, we will often find it advantageous 
to use the concrete and intuitive
terminology of information theory. We will for example often 
use terms such as
``signal" and ``bandwidth" where we mean ``field" and ``ultraviolet cut-off". 
For our purposes, 
the main advantage of the language of information theory will be that
this language contains several useful terms
which describe properties of bandlimited signals - and which by correspondence 
also describe properties of ultraviolet cut-off fields -  for which 
there does not seem to exist an established corresponding terminology in 
the language of quantum field theory. These will be terms such as 
``data transmission rate", ``noise", or ``signal reconstruction from samples".
We will introduce these terms as needed, and we will   
discuss their corresponding meaning in
quantum field theory.
\bn
Concerning the physics of superoscillations in ultraviolet regularized
quantum field theories, an interesting possibility has been discussed
in \cite{rosu,reznik}. These authors suggest that the existence
of superoscillations may resolve the transplanckian energies paradox of black
hole radiation:
\sn
Let us recall how this paradox 
arises for black hole radiation as derived
in the standard free field theory formalism.
One considers a Hawking photon which
is observed at asymptotic distance from the black hole 
and which has some typical energy $E$. 
The calculation of the redshift 
shows that the same photon, when it was still close to the horizon,
say at a Planckian distance, should have had 
a proper energy of the order of $E e^{\alpha M^2}$ where $\alpha$ is of order
one and where for a macroscopic black hole, in Planckian units, 
$M\approx 10^{40}$.
The paradox is that 
the assumptions which went
into the derivation of the existence of 
Hawking radiation (no interactions or backreactions) likely do not hold true at those far
transplanckian energies.  
See for example \cite{thooft1}-\cite{broutrev}. 
\sn
This, therefore, raises the question whether the phenomenon of Hawking 
radiation is dependent on assumptions about
the physics at transplanckian energies. In particular, 
a fundamental ultraviolet cut-off may well exist at a Planck- or string scale,
and the question arises whether
Hawking radiation is compatible with the existence of a natural ultraviolet cut-off. 
A number of studies have therefore investigated the problem of Hawking radiation
in the presence of various kinds 
of ultraviolet cut-off. For reviews, see e.g. \cite{jacobsen1} or  \cite{broutrev}.
An intuitive example is Unruh's consideration
of the dumb hole \cite{unruh1,unruh2}, which is
an acoustic analog of the black hole, where
a ``horizon" forms where a stationarily flowing fluid's velocity exceeds the velocity of 
the sound waves. 
\sn
The consensus in the literature appears to be that 
Hawking radiation is indeed a robust phenomenon - if only for example for
thermodynamical reasons. However, there does not seem to exist a consensus 
about how precisely the transplanckian frequencies
paradox is to be resolved. The recent work by Rosu \cite{rosu}
 and Reznik \cite{reznik} which we mentioned above
 aims at resolving the transplanckian energies paradox 
by employing superoscillations. Their main argument is that even fields with a
strict ultraviolet cut-off at some maximum frequency 
can still display arbitrarily high frequency oscillations,
indeed any transplanckian frequencies, 
 in some finite region, e.g. close to the horizon, namely if 
the field superoscillates.
\sn
In this context, as in all contexts where an argument is 
based on the phenomenon of
superoscillations, the argument can only be as good as our understanding of the
properties of superoscillations.
\bn
We will therefore address here three points concerning the general properties of 
superoscillations:
\bn Firstly, we will apply methods recently developed in
\cite{ak-shannon} to obtain 
exact results about the extent to which frequency limited functions can
superoscillate. Namely, we will show that among the functions with
frequency cut-off $\omega_{max}$ there always exist functions which pass
through any finite number of arbitrarily prespecified points. We will also show
that superoscillations cannot be prespecified on any continuous interval.
We can translate this result into the language of information theory:
\sn
The implication is that a 20KHz recording of a Beethoven
symphony cannot occur as part of a 1Hz bandlimited signal - but that
1 Hz bandlimited signals can indeed always be found which coincide with the
Beethoven symphony at arbitrarily many discrete points in time. 
\sn
Our result on the extent to which frequency limited functions can superoscillate
shows, in particular, that frequency limited
functions can indeed not be reliably characterized as varying
slower than their highest Fourier component.
This raises the problem of finding a reliable
characterization of the effect of frequency limitation on
the ``behavior" of functions.
\bn 
Therefore, secondly, we will show
that a reliable characterization of the effect of frequency limitation on
the behavior of functions is in terms of an uncertainty relation:
If a strictly frequency limited function, $\phi(t)$, superoscillating or not,
is sampled at the
so-called Nyquist rate, then the standard deviation $\Delta T$  of
its samples $\phi(t_n)$ is bounded from below by $\Delta T >
1/4\omega_{max}$. We will therefore 
conclude that a frequency limit is not a limit to
how quickly a function can vary, but that instead a frequency limit is
a limit to
 how much a function's Nyquist rate
samples can be peaked.
\bn
Thirdly, we will explain how this characterization
of frequency limited functions
generalizes to time-varying frequency limits $\omega_{max}(t)$.
We will apply these results to a recently
 developed generalized Shannon sampling theorem \cite{ak-shannon}.
\sn
Translated into the language of quantum field theory,
our results will show, for example, that 
a frequency cut-off can possess many of the 
advantages of lattice regularizations without
needing to break translation invariance. For example, the 
Shannon sampling theorem and its generalization implies 
that the number of degrees of freedom per
unit volume (we assume a euclidean formulation) is 
literally finite for
ultraviolet cut-off fields: the fields are fully 
determined if they are specified 
on any one of a family of equivalent lattices whose spacing is
determined by the ultraviolet-cut-off. The
family of lattices is
 covering the entire  continuous space 
so that translational invariance is not broken.
We will cover the general case where the density
of degrees of freedom is spatially varying.

%%%%%%%%%%%%%%%%%%%%%%%%%%%%%%%%%%%%%%%%%%%%%%%%%

\section{Examples of  Superoscillations}
Let us consider functions, $\phi(t)$, which are frequency limited
with a maximum frequency $\omega_{max}$,
i.e. which contain only plane waves up to this
frequency.
We can write such functions in the form
\be
\phi(t) ~=~ \int_{-\infty}^{+\infty}
 du~ r(u)~e^{i t \omega(u)}
\label{uno}
\ee
where $\omega(u)$ is a real-valued
function which obeys
\be
\vert \omega(u)\vert~ \le~ \omega_{max}
~~~~~~~~\mbox{for all} ~u \in \R~,
\ee
and where $r(u)$ is a complex-valued function.
\sn
Berry \cite{berry} gives the explicit example 
(among other examples)
\be
\omega(u) ~:=~ \frac{\omega_{max}}{1+u^2}
\label{due}
\ee
and
\be
r(u) ~:= ~ \frac{1}{\sqrt{2\pi\epsilon}}
~e^{-\frac{(u-i c)^2}{2\epsilon}},
\ee
where $\epsilon$ and $c$ are positive constants.
The claim is that for suitable choices
of $\epsilon$ and $c$
the resulting function $\phi(t)$ displays
superoscillations, i.e.
that in some interval it oscillates faster
than $\omega_{max}$.
\sn
There is a simple argument for why this
should be true. Berry reports this argument
 to be due to
Aharonov:
\sn
Namely, for sufficiently small $\epsilon$ the
function $r(u)$ should effectively
become a Gaussian approximation to a Dirac
$\delta$-function which is peaked around
the imaginary value $u=i c$. Therefore, the
factor $r(u)$
in Eq.\ref{uno} should effectively project
out the value of
the integrand at $u=ic$.
Due to Eq.\ref{due},
this value of $u$ corresponds to the frequency:
\be
\omega (ic)~ =~ \frac{\omega_{max}}{1-c^2}.
\ee
Clearly, for suitable choices of the parameter
$c$, this frequency
can be made arbitrarily larger than the bandwidth
$\omega_{max}$.
\sn
Thus, the situation is that on the one hand,
$\phi(t)$ certainly contains only frequencies up to $\omega_{max}$
because $\omega(u)\le \omega_{max}$ for all
\it real \rm values of $u$, and
the intergration in Eq.\ref{uno} is over real $u$ only.
On the other hand, for \it imaginary \rm values of $u$
the value of $\omega(u)$ can become much larger than
$\omega_{max}$ . Indeed,
 the behavior of  $r(u)$ indicates that
the integral should effectively be peaked around the imaginary
value $u=ic$. This suggests that, for choices of $c$ close enough
to $1$, in some interval the function $\phi(t)$ could display
superoscillations with frequencies around $\omega_{so}\approx
1/(1-c^2)>\omega_{max}$. \sn This intuitive argument for
superoscillations has been confirmed, in \cite{berry}, both by
asymptotic analysis and by numerical calculations. Berry also
explains in \cite{berry} that the price for a function to have
this type of a superoscillating period is that the function also
possesses a period with exponentially large amplitudes -
nevertheless, the whole function is square integrable. We remark that
another method for constructing examples of
superoscillations has been found in \cite{qiao}.

\section{To which extent are frequency limited functions able to superoscillate?}%*****
\subsection{Definitions}%**********************
Let us in the following  refer to frequency limited
functions $\phi(t)$ as
 ``signals" and to the variable $t$ as ``time".
\sn
 More precisely, we define the class of  \it signals \rm  $\phi$ with
\it bandwidth \rm $\omega_{max}$ as the Hilbert space of square integrable
functions on the interval
 $[-\omega_{max},\omega_{max}]$
in frequency space
\be
H_{\omega_{max}}
= L^2(-\omega_{max},\omega_{max})
\ee
with the usual scalar product:
\be
(\phi_1,\phi_2) ~=~
\int_{-\omega_{max}}^{\omega_{max}} d\omega~
\tilde{\phi}_1(\omega)^*~\tilde{\phi}_2(\omega)
\label{scproduct}
\ee
We then define the set $B_{\omega_{max}}$ of
\it strictly band\-limited signals \rm
with bandwidth $\omega_{max}$ as
the set of all functions $\tilde{\phi}(\omega)$
on frequency space
for which there
exists a $c(\phi)<\omega_{max}$ such that
\be
\tphi(\omega)~=~0~~~~~~~~\mbox{if}
~~~\vert\omega\vert > c(\phi)
\label{bwc}
\ee
and whose derivatives $d^n
 \tphi(\omega)/d\omega^n$ are
square integrable for all $n\in\N$.
\sn
Clearly, the strictly bandlimited signals are
dense in the Hilbert space of
 bandlimited signals $\Hi$:
\be
H_{\omega_{max}}~=~\overline{B_{\omega_{max}}}
\ee
\subsection{Proposition}%**************************
We claim that each Hilbert space of
  band\-limited signals $H_{\omega_{max}}$
 contains signals
such that the Fourier transform of
 $\tphi(\omega)$, i.e. the signal $\phi(t)$,
passes through any finite number of
arbitrarily prespecified points.
\sn
Explicitly, we can fix a value for the
 bandwidth, $\omega_{max}$. Then, we
choose $N$ arbitrary times $\{t_i\}_{i=1}^N$
and $N$
arbitrary amplitudes $\{a_i\}_{i=1}^N$. The
claim is that there always exist
signals of bandwidth  $\omega_{max}$
which obey:
\be
\phi(t_i)~=~ a_i ~~~~~\mbox{for all}~ ~i =1,2,...,N
\ee
In field theory language, we are claiming that for any choice
of an ultraviolet cut-off frequency 
there are fields which obey the cut-off and which 
at an arbitrary finite number of points in space
take arbitrary prespecified values. In particular,
since these points and the amplitude of the field at
these points can be chosen arbitrarily we claim that even fields 
which obey a cut-off can vary arbitrarily wildly
over any finite interval.

\subsection{Proof}%****************************
\label{fa}
Let us first outline the proof.
We will begin by considering
the simple symmetric operator
$T:~\phi(t)\rightarrow t\phi(t)$
on $\B$. Its self-adjoint extensions,
$T(\alpha)$, then yield a set of
 Hilbert bases $\{\ve_n(\alpha)\}$
of $H_{\omega_{max}}$ as their eigenbases.
The amplitudes of band\-limited
signals $\phi(t)$ can be written as
scalar products  with these
eigenvectors: $\phi(t)=(\ve,\phi)$. The
proof of the proposition
will  consist in showing that any finite
set $\{\ve_i\}_{i=1}^N$ of basis vectors
among all
eigenvectors of the self-adjoint
extensions is linearly independent.
\subsubsection*{The ``time operator"
$T$}%**********************************
We define the operator $T$ on the
domain $D_T :=B_{\omega_{max}}$ as the
operator which acts on
strictly band\-limited signals $\phi(t)$
by multiplication
with the time variable:
\be
T:~~ \phi(t) ~\rightarrow~ T\phi(t)~=~t~\phi(t)
\ee
The operator $T$ maps strictly band\-limited
functions into strictly band\-limited functions:
\be
T:~~B_{\omega_{max}}~\rightarrow~ B_{\omega_{max}}
\ee
This is because $T$ acts in the Fourier representation as
\be
T:~~ \tilde{\phi}(\omega) ~\rightarrow
~T\tilde{\phi}(\omega)=
-i\frac{d}{d\omega} \tilde{\phi}(\omega)
\ee
and, clearly, if $\tilde{\phi}(\omega)$ obeys
 the bandwidth condition, Eq.\ref{bwc}, so does
its derivative $\partial_\omega\tphi(\omega)$.
\sn
The elements $\phi \in D_T$ are strictly
 band\-limited and they therefore obey, in particular:
\be
\tilde{\phi}(-\omega) ~=~0~=~\tilde{\phi}(\omega)
\label{rbe}
\ee
Thus, for all $\phi\in \B$:
\be
\int_{-\omega_{max}}^{\omega_{max}}
d\omega~{\tphi}_1(\omega)^*
(-i\partial_\omega)\tphi_2(\omega)
~=~
\int_{-\omega_{max}}^{\omega_{max}} d\omega~
\left((-i\partial_\omega)
{\tphi}_1\right)^*   \tphi_2(\omega)
\ee
Consequently,
\be
(\phi, T \phi) ~=~ (T \phi, \phi) ~=~ (\phi,
T\phi)^*~~~~~~~~\forall~\phi \in \B
\ee
and therefore,
\be
(\phi,T\phi) ~\in~\R ~~~~~~~~\forall~\phi\in \B
\ee
which means that $T$ is a symmetric operator.
\sn
Nevertheless, $T$ is not self-adjoint. Indeed,
$T$ possesses
no  (normalizable nor nonnormalizable) eigenvectors.
This is because the only candidates for eigenvectors,
namely the plane waves $e^{2\pi i t\omega}$
do not obey Eqs.\ref{bwc},\ref{rbe}. Thus, the plane
waves
are not strictly band\-limited and therefore
they are not in the domain $D_T=\B$ of $T$. On the
 other hand, while the plane waves are
not strictly band\-limited, they are nevertheless
band\-limited, i.e.
they are elements of the Hilbert space $\Hi$.
Indeed, the domain of $T$ can be suitably enlarged
to yield a whole family of self-adjoint extensions
of $T$, each with a
discrete subset of the plane waves as an eigenbasis.
We will derive these self-adjoint extensions below.
For a standard reference on the functional analysis
 of self-adjoint extensions, see e.g. \cite{AG}. 
\subsubsection*{The self-adjoint extensions
$T(\alpha)$ of $T$, and their eigenbases} %**********
%*******************************************************
There exists a $U(1)$- family of
self-adjoint extensions $T(\alpha)$ of $T$:
\sn
The self-adjoint operator $T(\alpha)$ is obtained
by enlarging the domain of $T$ by signals, $\phi$,
which obey the boundary condition:
\be
\tphi(-\omega) ~=~e^{i\alpha}~\tphi(\omega)
\label{bccc}
\ee
To be precise: We first close the operator $T$.
Then, the domain $D_{T^*}$ of $T^*$ consists of
all those
signals $\phi\in H$ for which also
$-i\partial_\omega \tphi(\omega) \in\Hi$.
The signals  $\phi\in D_{T^*}$ are not
required to obey any boundary conditions.
Thus, all plane waves are eigenvectors of $T^*$.
Note that while some plane waves are orthogonal,
 most are not.
This is consistent because $T^*$ is
not a symmetric operator: due to the lack of
boundary conditions in its domain, $T^*$ also has complex
expectation values. Any self-adjoint extension
$T(\alpha)$ of $T$
is a restriction of $T^*$ by imposing a boundary
condition of the form of Eq.\ref{bccc}:
\be
D_{T(\alpha)} ~=~\{\phi\in D_{T^*}\vert  ~
\tphi(-\omega) ~=~e^{i\alpha}~\tphi(\omega)\}.
\ee
For each choice of a phase $e^{i\alpha}$ we obtain
 an operator $T(\alpha)$ which is self-adjoint
and diagonalizable. Its orthonormal
eigenvectors,
$\{\ve_n^{(\alpha)}\}_{n=-\infty}^{+\infty}$,
obeying
\be
T(\alpha) \ve_n(\alpha) = t_n(\alpha) \ve_n(\alpha),
\ee
form a Hilbert basis for $\Hi$. In
frequency space, they are
the plane waves
\be
\tilde{\ve}_n^{(\alpha)}(\omega) ~=~ \frac{e^{2\pi
 i t_n(\alpha) \omega}}{\sqrt{2\omega_{max}}}
\ee
which correspond to the $T(\alpha)$-eigenvalues:
\be
t_n(\alpha) ~=~ \frac{n}{2\omega_{max}}
-\frac{\alpha}{4\pi\omega_{max}},~~~~~~~n\in \Z
\label{evs}
\ee
As mentioned before, each eigenvector of a self-adjoint
 extension is also an eigenvector of
$T^*$, the adjoint of $T$:
\be
T^*\ve_n(\alpha) ~=~t_n(\alpha)
~\ve_n(\alpha)~~~~~~~~~~\forall~n,\alpha
\label{ts}
\ee
The eigenvalues of $T^*$, i.e. the eigenvalues
of all the extensions $T(\alpha)$, together,
cover the real line
exactly once, i.e. for each $t\in\R$ there
exists exactly one $e^{i\alpha}$
and one $n$ such that $t=t_n(\alpha)$.
We will therefore occasionally write simply
$\ve$ for $\ve_n(\alpha)$. In this notation,
Eq.\ref{ts} reads:
\be
T^*\ve ~=~t ~\ve
\ee
Using the scalar product, Eq.\ref{scproduct},
the signal $\phi(t)$, i.e. the Fourier transform
of the function $\tphi(\omega)$,
can then be written simply as:
\be
\phi(t)~=~ (\ve,\phi)
\label{re}
\ee
Thus, the signal as a time-dependent
function $\phi(t)$ is the
expansion of the abstract signal $\phi$
in an overcomplete set of vectors, namely
in all the eigenbases of the family of
operators $T(\alpha)$.
\sn
As an immediate consequence we recover
the Shannon sampling theorem:
\subsubsection*{The Shannon sampling
theorem, and its translation into field theory terminology}%*********************************
The Shannon sampling theorem states that
 if the amplitudes of a strictly
band\-limited signal $\phi(t)$
are known at discrete points in
time with spacing
\be
t_{n+1}-t_n=1/2\omega_{max}
\ee
which is
the so-called Nyquist rate, then
the signal $\phi(t)$
can already be calculated for all $t$:
\sn
Namely, let us fix one $\alpha$. Then, to
know the values $\phi(t_n(\alpha))$
of the function $\phi(t)$ at the discrete
set of eigenvalues $t_n(\alpha)$ (whose spacing,
 from Eq.\ref{evs}, is $1/2\omega_{max}$),
is to know the coefficients of the vector $\phi$
in the
Hilbert basis $\{\ve_n(\alpha)\}$. Thus, $\phi$
is fully determined
as a vector in the Hilbert space $\Hi$. Therefore,
its coefficients can be calculated in any
arbitrary Hilbert basis. Thus, in particular, the
 values of
$\phi(t)=(\ve,\phi)$ can be calculated for all $t$:
\be
\phi(t)~=~ \sum_{n=-\infty}^\infty
~(\ve,\ve_n(\alpha))~\phi(t_n(\alpha))
\label{shannon}
\ee
Clearly, Eq.\ref{shannon} is obtained simply
by inserting the resolution of the
identity $1=\sum_{n=-\infty}^{\infty}
\ve_n(\alpha)\otimes \ve^*_n(\alpha)$ on the
 RHS of Eq.\ref{re}.
We note that
while for each fixed $\alpha$ the set of vectors
 $\{\ve_n(\alpha)\}$ forms an orthonormal
Hilbert basis in $H$, the basis vectors belonging
 to different
self-adjoint extensions are not orthogonal:
\be
(\ve_n(\alpha),\ve_m(\alpha^\prime))\neq 0~~~~~~~\mbox{ for}~~
\alpha \neq \alpha^\prime.
\ee
In the sampling formula Eq.\ref{shannon} we need
this scalar product, i.e. $(\ve,\ve_n(\alpha))$,
and it is easily calculated for all values of
 the arguments:
\begin{eqnarray}
(\ve,\ve^\prime) & = &
 \int_{-\omega_{max}}^{\omega_{max}}d\omega~
\frac{e^{2\pi i(t-t^\prime)\omega}}{2\omega_{max}}
 \nonumber
\\
\\
  & = & \frac{
\sin( 2\pi(t-t^\prime)\omega_{max})
}{2\pi(t-t^\prime)\omega_{max}}~
\nonumber
\end{eqnarray}
Note that the sampling kernel $(\ve,\ve^\prime)$
is real and continuous
which means that we describe real, continuous
(in fact, entire) signals $\phi(t)^*=\phi(t)$, which
 would not be the case
for other choices of the phases of the
eigenvectors $\ve$.
\bn
The Shannon sampling theorem has an interesting translation into
the language of field theory: 
Consider first fields, say scalar fields, without an ultraviolet cut-off. These
fields possess at each point in space one degree of freedom: the amplitude.
Thus, the field possesses an infinite number of degrees of freedom
per unit volume. 
\sn 
The Shannon sampling theorem shows that 
an ultraviolet cut-off field is already determined everywhere
if it is known only on any one of a family of discrete lattices. 
In other words, fields which are ultraviolet cut-off, in the original sense
of a frequency cut-off, are continuous fields, which can however
be represented without loss of information on certain discrete lattices. 
This also means
that for ultraviolet cut-off fields the number of degrees of freedom of 
per unit volume is literally finite: it is given by the number of 
sampling points needed per unit volume in order to be able
to reconstruct the field everywhere. The field theoretic meaning of
the information theory term ``Nyquist rate" is the spatial density of 
the degrees of freedom of fields. We will later discuss
a generalization of the Shannon sampling theorem for classes of
signals whose Nyquist rate is time-varying. This theorem will 
translate into the statement that these signals with time-varying
bandwidth correspond to fields whose spatial density of degrees of freedom
is spatially varying. These are continuous fields which are representable
without loss of information on families of lattices whose minimum spacing is 
spatially varying.

\subsubsection*{Superoscillations}%***
We can now prove that for every bandwidth $\omega_{max}$
 there  always exist
band\-limited signals $\phi\in \Hi$, which pass
through any finite number of prespecified points.
\sn
To this end we choose $N$ arbitrary distinct
times $t_1,...,t_N$
and $N$ amplitudes $a_1,...,a_N$. We
must show that for each such choice and for
each bandwidth $\omega_{max}$ there exist
band\-limited signals $\phi\in \Hi$ which
pass at the times $t_i$
through the values $a_i$:
\be
\phi(t_i)~ =~(\ve_i,\phi)~=~ a_i
~~~~~~~~~\forall ~i=1,...,N
\label{cono}
\ee
We recall that the eigenbases of the self-adjoint
 extensions $T(\alpha)$ of $T$ each
yield a resolution of the identity:
\be
1~=~\sum_{n=-\infty}^{+\infty} \ve_n(\alpha)\otimes
 \ve^*_n(\alpha)
\ee
Inserting one of these resolutions of the identity
into Eq.\ref{cono} we obtain
an explicit inhomogeneous system of linear equations:
\be
\sum_{n=-\infty}^{+\infty}
 ~(\ve_i,\ve_n(\alpha))~(\ve_n(\alpha),\phi)
  ~=~a_i~~~~~~~~~~\forall ~i=1,...N
\label{inho}.
\ee
Solutions to Eq.\ref{inho} exist, i.e. there
are band\-limited signals
which go through all the specified points, exactly
 if the matrix $(\ve_i,\ve_n(\alpha))$ is of
full rank
\be
\mbox{rank}\left((\ve_i,\ve_n(\alpha))_{i=1,~~
n=-\infty}^{i=N,~~n=+\infty}\right)~=~N,
\label{rank}
\ee
which is the case exactly if  the set of vectors
$\{\ve_i\}$  is linearly independent.
\sn
In order to prove that indeed every finite set of
distinct eigenvectors $\ve_i$ of $T^*$ is
linearly independent, let us now assume the opposite.
Namely, let us assume
that there does exist a set of $N$ eigenvectors
$\ve_i$ of  $T^*$,
and complex coefficients $\lambda_i$ which are
not all zero, such that:
\be
\sum_{i=1}^N~\lambda_i ~\ve_i~=~0
\label{linde}
\ee
Since the sum is a finite sum, we can repeatedly
 apply $T^*$ to
Eq.\ref{linde}, to obtain:
\be
\sum_{i=1}^N~\lambda_i ~t^n_i ~\ve_i~=~0
~~~~~~\forall~ n\in \N
\ee
The first $N$ equations yield:
\be
\left( \matrix{1 & 1 & 1 & ... & 1\cr
t_1 & t_2  & t_3 & ... & t_N\cr \cr
  & &   \vdots & &  \cr \cr
t_1^{N-1} & t_2^{N-1} & t_3^{N-1} & ... & t_N^{N-1}}
\right) ~~\left( \matrix{\lambda_1 ~ \ve_1\cr
\lambda_2 \ve_2 \cr \lambda_3 \ve_3\cr \cr
\vdots \cr \cr \lambda_N \ve_N}\right)
~~=~~ 0
\label{vmm}
\ee
This $N\times N$ matrix is a Vandermonde matrix
 and its determinant is known to
take the form:
\be
\left\vert \matrix{1 & 1 & 1 & ... & 1\cr
t_1 & t_2  & t_3 & ... & t_N\cr \cr
  & &   \vdots & &  \cr \cr
t_1^{N-1} & t_2^{N-1} & t_3^{N-1} & ... & t_N^{N-1}}
\right\vert
~=~ \prod_{1\le j< k \le N}~(t_k-t_j)
\ee
In particular, the determinant does not vanish,
since the $\ve_i$ are by assumption distinct, i.e.
$t_k\neq t_j$ for all $k\neq j$. Thus, the
Vandermonde matrix has an inverse. Multiplying this
inverse from the left onto Eq.\ref{vmm}
we obtain that $\lambda_i \ve_i=0, ~\forall i=1,...,N$,
i.e. we can conclude
that $\lambda_i=0 ~ \forall~i=1,...,N$.
\sn
Therefore, any finite set of distinct eigenvectors
 $\ve$ of $T^*$ is indeed linearly independent and
consequently Eq.\ref{rank} is obeyed.
\sn
Thus, for any arbitrarily chosen bandwidth $\omega_{max}$,
 there are indeed
signals $\phi\in \Hi$ which pass
through any finite number of arbitrarily
prespecified points.
\subsection{Beethoven at 1Hz ?}%************
Let us now address the question whether a recording of a Beethoven
symphony could indeed appear as part of a 1Hz bandlimited signal.
Correspondingly, the question is whether
 fields on a space with this particular ultraviolet cut-off
are free to take prespecified values on a finite interval. \sn
More explicitly, let us ask for example whether it is possible to
take say $5000$ seconds of a 20 KHz recording
 of a Beethoven symphony and to append
a suitable function before and suitable function
after the symphony, so that the whole signal
ranging from time
$t=-\infty$ to $t=+\infty$ is a 1Hz band\-limited
 signal.
 \sn If the question is posed in this form, the answer is
no. To see this, we recall that bandlimited functions are always
entire functions. Entire functions are Taylor expandable
everywhere,
 and with infinite radius of convergence.
Thus if an entire function $\phi(t)$ is known on even
 a tiny interval $[t_i,t_f]$ of the time axis, then
we can calculate at a point $t_0\in [t_i,t_f]$ in
that interval all derivatives $d^n/dt^n\phi(t_0)$.
This yields a Taylor series expansion of $\phi(t)$
around the time $t_0$
with infinite radius of convergence.
Thus, if a band\-limited function is
known on any finite interval then it is already
determined everywhere.
\sn
One consequence is that a band\-limited
signal cannot
vanish on any finite interval, since
this would mean that it vanishes everywhere.
Thus, for example, if the original signal of the
 Beethoven recording is truly 20KHz
band\-limited, then it is an entire function and
therefore it does not vanish on any finite
interval between $t=-\infty$ and $t=+\infty$.
On the other hand, we are only interested in an
interval of length about $5000s$.
Now the question is whether these 5000 seconds of
 the 20KHz band\-limited
recording can occur as a superoscillating
period of a signal which is band\-limited, say by
1Hz. The answer is negative because this
1Hz bandlimited signal, if existing, would also be
entire - but clearly two
entire functions which coincide on a finite interval
 coincide everywhere.
\sn
It is therefore not possible to arbitrarily
prespecify the exact values of a 1Hz
band\-limited signal on any finite interval. 
We are left with the question whether
there are topologies with respect to which
approximations may converge. On the other
hand, it is clear that if we wish to prespecify 
precise values of the signal then the
most that may be possible is to arbitrarily
prespecify the values of a 1Hz band\-limited signal
 at arbitrary
discrete times. This would mean, for example, that
one can find 1Hz band\-limited signals which
coincide with the 20KHz Beethoven recording at
arbitrarily many discrete points
 in time. That it is indeed possible to prespecify the 
signals' values at an arbitrary finite number 
of discrete points in time is what we
  proved in the previous section.
\subsection{Superoscillations for
data compression?}%****
As is well-known, the bandwidth
of a communication channel limits its
 maximal data transmission rate.
We have just seen, however, that signals
with
fixed bandwidth can superoscillate and
exhibit for example arbitrarily fine
ripples and arbitrarily sharp spikes.
This suggests that it should be possible to encode and transmit 
an arbitrarily large amount
of information in an arbitrarily short time interval of
a 1Hz bandlimited signal - because there is always a
1Hz bandlimited signal which passes through any
number of arbitrarily prespecified points.
\sn
Thus, this raises the question whether
superoscillations are able to circumvent the
 bandwidth limitations of communication channels -
and whether, as Berry suggested,
superoscillations may for example be used for
data compression.
\sn
Here, we need to recall that the bandwidth alone
does not fix the maximal data transmission rate.
It is known that, in the absence of noise, every channel -
 with any arbitrary
bandwidth - can carry an infinite amount of information in
any arbitrarily short amount of time. 
\sn
In practise, every channel has noise and this prevents us from 
measuring the signal to ideal precision. Essentially, the effect of the
noise is that only a finite number of amplitude levels can be resolved.
Now if the information
is encoded in $V$ different amplitude levels
(i.e. binary would be two levels, $V=2$),
then the maximum baud rate $b$ in bits$/$second is
\be
 b~=~ 2\omega_{max} ~\ln_2V.
 \label{plain}
\ee
This follows immediately from the Shannon sampling theorem:
Each amplitude measurement yields one out of $V$ possible outcomes,
i.e. each measurement yields $\ln_2V$ bits of information.
This yields Eq.\ref{plain} because by the Shannon theorem
we need to measure only $2 \omega_{max}$ samples per second to capture all
of the signal.
We only remark here that for the example of  white noise the maximal
data transmission rate can be expressed directly in the signal to noise ratio $S/N$:
\be
b_{noise}~=~\omega_{max}~\ln_2\left(1+\frac{S}{N}\right)
\label{noise} \ee For the precise definitions and the proof, see
e.g. the classic text by Shannon, \cite{shannon-mtc}.
\sn
For us interesting here is that Eqs.\ref{plain},\ref{noise} show
that indeed even in the presence of noise the
data transmission rate can be made arbitrarily large, for any
fixed bandwidth - though at a cost. The price to be paid is
that in order to increase the baud rate to bandwidth ratio 
the maximal signal amplitude must be improved exponentially
as compared to the resolution of the amplitude, or more precisely as
compared to the noise level. 
\sn 
Let us consider the implications for superoscillations.
Superoscillations, in spite of their peculiar behavior,
do obey the bandlimit, $\omega_{max}$. Therefore,
superoscillations cannot violate the limits on the baud rate in Eqs.\ref{plain},
\ref{noise}. Indeed, conversely, from the validity of the limits
on the baud rate we can deduce properties of superoscillations:
If large amounts of information are to be sent over a
low bandwidth channel, e.g. by employing superoscillations, 
this necessitates an exponentially large dynamical range of the
superoscillating signal. Indeed, 
Berry conjectured that superoscillations necessarily
occur with exponentially large dynamical ranges.
\mn 
This is essentially the same as saying that it is difficult to 
stabilize superoscillations under perturbations:
\sn We showed
that it is not possible to prespecify superoscillations on any \it
continuous time interval. \rm For example, there is no 1Hz
bandlimited function which coincides with a symphony's recording
on a continuous interval of say 5000 seconds. On the other hand, we
showed that it is possible to prespecify
 superoscillations at any number
of \it discrete points in time. \rm For example,
there do exist 1Hz bandlimited functions which
 coincide with the
20KHz bandlimited Beethoven recording at $10^{1000}$
 points in time
during the 5000 seconds of the recording. \sn Thus, a 1Hz
bandlimited function which coincides with a symphony's recording
at $10^{1000}$ points on a 5000s interval, can only be 1Hz
bandlimited because of very fine-tuned cancellations in the calculation
of its Fourier spectrum - cancellations which depend on 
small details of the function. We therefore
conclude that tiny perturbations of  such a
1Hz bandlimited superoscillating function are able to induce very
high frequency components. 
Thus, superoscillations are in this
sense unstable, and they are therefore likely to be difficult
to make practical use of in imperfect communication channels. 
\sn
On the other hand, as the reverse side of the
coin, important phenomena in signal
processing are instabilities in the
reconstruction procedures of signals which
are oversampled, i.e. which are sampled at 
a rate higher than the Nyquist rate. The
instabilities in the reconstruction
arise because small
imprecisions in the measurement
of the then overdetermined samples of an ordinary 
(i.e. in general non-superoscillating) signal 
can lead to the reconstruction of a deviant signal 
which, in our terminology, possesses 
superoscillations. 
This connection was pointed out already by 
Berry \cite{berry}, quoting I. Daubechi. For a general reference on
oversampling see e.g. \cite{marks}.
\sn
In terms of models of
fields at the Planck scale, the instabilities
of superoscillations  suggest that
in ultraviolet cut-off quantum  field theories
the intercation of particles whose fields superoscillate 
could easily destroy their 
superoscillations. In concrete cases this effect is likely to depend 
on the type interactions
of the field theory that one considers. Studies in this direction
 could be worth pursuing since these instabilities could
 have implications  for example
for the viability of the Rosu-Reznik approach to
superoscillations in black hole radiation
when treated within a framework of interacting fields.
\section{A strict bandlimit is a lower bound
to how much the samples of a signal can be peaked - an uncertainty relation
for ultraviolet cut-off fields}%***************

\subsection{The minimum standard deviation}
The existence of superoscillations shows that
band\-limited functions cannot be characterized
reliably as varying at most as fast as their
highest Fourier component.
Indeed, we have just proved that
for any  fixed bandwidth there always exist functions
which possess
arbitrarily fine ripples and arbitrarily sharp
spikes.
Let us therefore look for a better, i.e. for a
 reliable characterization of
the effect of band\-limitation on the behavior
of functions.
\mn
Our proposition is that, while
a bandlimit does not imply a bound on how much
band\-limited signals can
\it locally \rm be peaked,
a bandlimit does imply a bound to how much
strictly band\-limited signals can be peaked
 \it globally. \rm 
Equivalently, our proposition is that while an ultraviolet cut-off
does not imply a bound to how much the fields can be peaked 
locally in space, the cut-off does imply a lower bound to how much
the fields can be peaked globally. 
\sn
Our motivation derives from the Heisenberg
uncertainty principle:
If we read $T$ as the momentum operator of a
particle in a (one-dimensional) box
then, because the position uncertainty is bounded
 from above by the size of the box,
 we expect the
momentum uncertainty (here $\dT$)
 to be bounded from below.
\sn
To be precise, consider a normalized,
 strictly band\-limited signal
$\phi\in \B$. Then,
\be
\overline{T}(\phi)~:=~(\phi, T\phi)
\label{expe}
\ee
is the $T$-expectation value, or the time-mean or the
``center of mass" of the signal $\phi$
on the time axis.
A measure of how much the signal is overall
 peaked around this time
is the formal standard deviation:
\be
\Delta T(\phi) ~:= ~ \sqrt{(\phi,
\left(T-\overline{T}(\phi)\right)^2 \phi)}
\label{fsd}
\ee
We note that both, $\oT$ and $\dT$ are not
sensitive to local features of $\phi(t)$, such as
fine ripples and sharp spikes. Instead, being
the first and second moment of $T$, the time
$\oT$ is simply the signal's global
average position on the time axis and $\dT$ is
the global spread of the signal around that position.
\sn
Our claim is that strictly band\-limited
signals, $\phi\in \B$, are
always globally spread by at least a certain
 minimum amount:
\be
\dT~> ~\frac{1}{4\omega_{max}}~~~~~~\mbox{for
 all}~~~~~\phi\in \B
\label{ucrn}
\ee
In field theory language, our claim is that there exists a formal 
finite minimum uncertainty
in position for ultraviolet cut-off  fields. 
\subsection{The minimum standard deviation as a
property of the Nyquist rate samples}%********************
Let us now rewrite $\oT$ and $\dT$  as explicit
expressions in the signals $\phi(t)$ as functions
of time. To this end, we can use any one of the
resolutions of the identity
$1=\sum_{n=-\infty}^{+\infty} \ve_n(\alpha)\otimes
\ve^*_n(\alpha)$ which are induced
by the self-adjoint extensions $T(\alpha)$ of $T$.
Inserting one of the resolutions of the identity
into Eq.\ref{expe} we obtain, restricting attention
to signals $\phi\in \B$
which are real,  $\phi(t)^*=\phi(t)$:
\be
\oT ~=~ \sum_{n=-\infty}^{\infty}\phi(t_n(\alpha))^2~
t_n(\alpha),
~~~~~~~(\mbox{independently of } \alpha)
\ee
Thus, $\oT$ is the ``mean" of the
discrete set of samples of the signal, when
sampled on one of the time-lattices of Eq.\ref {evs},
 i.e., $\oT$ is the time
around which the discrete samples of the signal
 $\phi$ are centered.
Indeed, for each set of samples taken at the
Nyquist rate (i.e. for each time lattice
corresponding to some fixed $\alpha$),
the time $\oT$ around which the samples are
centered is the same.
This is because
in order to calculate $\oT$ from Eq.\ref{fsd} we
 can equivalently
use any one of the
resolutions of the identity $1=\sum_{n=-\infty}^{+\infty}
 \ve_n(\alpha)\otimes \ve^*_n(\alpha)$.
\sn
Similarly, we obtain an explicit expression for
 how much the  samples
are spread around the value $\oT$ by inserting
a resolution of the identity into the expression
for the standard deviation,
Eq.\ref{fsd}:
\be
 \Delta T(\phi)~=~ \sqrt{\sum_{n=-\infty}^\infty
  \phi(t_n(\alpha))^2 ~\left(
t_n(\alpha) ~-~\oT\right)^2 }
\label{evalu}
\ee
Again, also the standard deviation does not depend
 on which sampling lattice $\{t_n(\alpha)\}$
has been chosen. We remark that, clearly, not only
the mean and standard deviation, but indeed
also all higher moments of a band\-limited signal's
Nyquist rate samples are independent of the
choice of the lattice of sampling times. We can
therefore refer to the mean, the standard deviation
and to the higher moments of a signal $\phi$ without
 needing to specify the choice of a
sampling lattice.
\sn
On the other hand, let us emphasize that
the values of $\oT$ and $\Delta T(\phi)$ are not
the usual mean and standard deviation of a continuous
curve as conventionally calculated
in terms of integrals rather than sums. Instead,
 while the strictly band\-limited signals
are of course continuous, $\oT$ and $\dT$ are the
mean and the standard deviation
of their discrete Nyquist rate samples.
\sn
Our proposition of above, i.e.  Eq.\ref{ucrn}, if
expressed explicitly in terms of the strictly
band\-limited signal's Nyquist rate samples,
is therefore that the standard deviation $\dT$
of these samples is bounded from below by
$1/4\omega_{max}$.
\subsection{Calculation of the maximally
 peaked signals$/$fields}
In order to prove the lower bound on the
standard deviation expressed in Eq.\ref{ucrn},
let us now explicitly solve the variational
problem of
finding signals $\phi$ which minimize
$\dT$. To this end, we minimize $(\phi, T^2 \phi)$
while enforcing the constraints
$(\phi, T\phi)=t$ and $(\phi,\phi)=1$.
\sn
We work in frequency space, where $T$ acts
on the strictly band\-limited signals
as the symmetric operator $T= -i ~d/d\omega$.
\sn
Introducing Lagrange multipliers $k_1,k_2$,
the functional to be minimized reads:
\be
S[\phi]~:=~\int_{-\omega_{max}}^{\omega_{max}}
d\omega
~\left\{-(\partial_\omega{\tilde{\phi}}^*)
(\partial_\omega{\tilde{\phi}}) +
k_1 ({\tilde{\phi}}^*{\tilde{\phi}} - c_1) +
k_2 (-i {\tilde{\phi}}^*\partial_\omega{
\tilde{\phi}}-c_2)\right\},
\ee
Setting $\delta S[\phi]/\delta\phi=0$ yields
the Euler-Lagrange equation:
\be
\partial_\omega^2 {\tilde{\phi}} + k_1 {\tilde{\phi}}
 -i\partial_\omega{\tilde{\phi}} ~=~0
\ee
Imposing the boundary condition, Eq.\ref{rbe}, which
 is obeyed
by all strictly band\-limited signals,
we obtain exactly one (up to phase)
normalized solution $\Phi_{\overline{T}}$ for
each value of  the mean $\overline{T}$:
\be
{\tilde{\Phi}}_{\overline{T}}(\omega)~=~
\frac{1}{\sqrt{2\pi \omega_{max}}}
~\cos\left(\frac{\pi~ \omega}{2
~\omega_{max}}\right)~e^{2 \pi
i \overline{T} \omega}
\ee
The standard deviations, $\Delta
T(\Phi_{\overline{T}})$, of these solutions
are straightforward to calculate
in Fourier space, to obtain:
\be
\Delta T(\phi_t) ~=~ \frac{1}{4\omega_{max}}~~
~~~\mbox{for all $t$}
\ee
Since the signals ${\tilde{\Phi}}_{\overline{T}}(\omega)$
 which minimize $\Delta T$
are not themselves strictly band\-limited - they do not
 obey Eq.\ref{bwc} -
we can conclude that all strictly band\-limited signals,
or ultraviolet cut-off fields, obey the strict bound
given in Eq.\ref{ucrn}.
\section{Generalization to time-varying bandwidths - or spatially varying
ultraviolet cut-offs}%****
\subsection{Superoscillations and the concept of
 time-varying bandwidth}
Intuitively, it is clear that the
bandwidths of signals can vary with time.
One might therefore expect to be able to define
the time-varying bandwidth of signals for example
in terms of the highest
frequency components which they contain in intervals
centered around different times.
This approach encounters difficulties, however, due
to the existence of superoscillations:
\sn
We recall that a signal $\phi(t)$ obeys a
\it constant \rm  bandlimit $\omega_{max}$ if
its Fourier transform
\be
\tphi(\omega)
=(2\pi)^{-1/2}\int_{-\infty}^{+\infty} dt ~\phi(t)
\exp(2\pi i \omega t)
\label{fourier}
\ee
has support only in the interval
$[-\omega_{max},\omega_{max}]$.
The integration in Eq.\ref{fourier} ranges over
the entire time axis. This means
that the bandlimit is a \it global \rm  property
of the signal.
\sn
If it were true that band\-limited signals could
nowhere vary faster than their highest
frequency component then
this would mean that the band\-width is
also a \it local  \rm property of the signal.
Namely, one might then expect that
if we consider the same signal on some finite
interval, $[t_i,t_f]$, and if we
calculate its Fourier
expansion on that interval then we will find that
 its Fourier coefficients are nonzero only for
frequencies smaller or equal than $\omega_{max}$.
If so, we could indeed define
 time-varying band\-widths as
time-varying upper limits on the
local frequency content, as indicated above.
Indeed, in practise,
windowed Fourier transforms and in particular 
the more sophisticated Wigner transforms or wavelet decompositions,
are generally very useful \cite{cohen,daubechies,klauder}.
\sn
However,  the
existence of superoscillations shows
that any \it local \rm definition of a
time-varying bandwidth  must contain counterintuitive features:
This is because whatever the overall bandwidth $\omega_{max}$,
there are always signals with this bandwidth which
superoscillate in any given interval
$[t_i,t_f]$. In practise, of course, 
strongly superoscillating signals will rarely occur because they
are very fine-tuned. But their existence shows that there do exist 
low bandwidth signals
 which locally possess arbitrarily
high frequency components 
- where ``local frequency components" are defined e.g. by
windowed Fourier transforms - 
in any finite length interval.
\sn
In field theory terminology this means that even if a field is
varying wildly in some spatial region, this does not imply that the field
necessarily possesses a large cut-off frequency 
or, equivalently, that it possesses a high density of degrees of freedom.
Instead, even at small cut-off frequencies there are fields
which locally display fast oscillations. Even these superoscillating 
fields are 
fully determined everywhere (by the sampling theorem)
if known only on any 
one of the family of lattices whose lattice spacing is
as large as is consistent with the ultraviolet cut-off. 

\subsection{The time-varying bandwidth as a
limit to how much the samples of signals
 can be peaked around different times}%**
We saw that a finite bandwidth does not impose
 a limit to how much
signals can be \it locally \rm peaked around
 say a time $t$. However, we also saw that
a finite  bandwidth does impose
a limit $\Delta T_{min}$ to how much the
signals can be \it globally \rm
peaked, around any time $t$. Indeed, this
characterization of
 the effect of bandlimitation naturally generalizes
to time-varying bandwidths:
Namely, the limit to how much signals
can be peaked may in general
depend on the time $t$
around which they are peaked:
\sn
We found that a constant bandwidth can be
understood as a minimum standard deviation
 of the signals' Nyquist rate samples:
If a strictly band\-limited signal $\phi\in \B$
is centered around a time $t= \oT$, then its
standard deviation around the time $t$ is always
 bounded from below by
the uncertainty relation $\dT >1/4\omega_{max}$.
\sn
We were then only discussing the case of constant
bandwidth. Accordingly, we
found that the standard deviation of signals
$\phi\in\B$ which are centered
around a time $t_1$ obey the same lower bound
$1/4\omega_{max}$
as do signals $\phi^\prime\in\B$ which are
centered around some other time $t_2$.
\sn
This suggests to try to define the notion of
time-varying bandwidth in such
a way that
a class of strictly band\-limited signals with a
time-varying bandwidth is simply a class of
signals for which the
minimum standard deviation $\Delta T_{min}$
depends on the time $t$ around
which the signals are centered. This would
mean that the uncertainty relation Eq.\ref{ucrn}
becomes time dependent:
\be
\Delta T(\phi) ~> ~ \Delta T_{min}(\oT)
\ee
Correspondingly, we would expect the Nyquist
rate to be time-varying.
\sn
To this end, let us recall the
functional analytic structure of  the Hilbert
 space of band\-limited signals
which we discussed in Sec.\ref{fa}: The operator
$T$ is a
simple symmetric operator with deficiency
indices $(1,1)$, whose self-adjoint extensions
have purely discrete and \it equidistant \rm spectra.
\sn
Indeed, the theory of simple symmetric operators
with deficiency indices $(1,1)$, whose self-adjoint
 extensions have discrete \it but not necessarily
equidistant \rm
spectra, has been shown to yield a generalized
Shannon sampling theorem in \cite{ak-shannon},
and it is indeed exactly the theory of time-varying
 bandwidths in the sense
which we just indicated. For example, the nonequidistant
spectra yield time-varying Nyquist rates.
The time-varying Nyquist rate
can be calculated from
the time-varying
minimum standard deviation $\Delta T_{min}(t)$
 and vice versa. This is worked out in detail in
\cite{ak-shannon2}.
\bn
In terms of field theory, the time-varying bandwidth means,
as we mentioned already, a spatially varying ultraviolet cut-off.
We remark that this is a nontrivial generalization of the concept of
frequency (or energy-momentum) cut-off in field theories.
An ordinary energy-momentum cut-off affects fields globally, i.e.
the cut-off scale is the same everywhere in space. 
But we may ask:
how could an energy momentum cut-off be implemented such that the
cut-off-scale is spatially varying?
 For example, the cut-off scale may be dynamically
generated, e.g. through an interplay of gravity 
and ordinary forces. In such a 
scenario, the actual cut-off scale may be dynamic and spatially 
varying, e.g. determined by the value of some field. 
In our approach to defining spatially
 varying cut-offs the cut-off is understood
as a formal  minimum uncertainty or
 standard deviation in position. For constant
bandwidths this is an equivalent definition 
to the usual definition as a frequency cut-off.
We then found that the notion of formal minimum 
position uncertainty generalizes `naturally'
to the situation where the value of the
 formal minimum position uncertainty 
depends on the position, i.e. on the formal 
position expectation value of the field.
As we will discuss in the last section,
 there is in fact very little arbitrariness
in the definition of these short-distance structures.

\section{Outlook}%*******************************
Functions with a bounded Fourier spectrum 
appear in numerous contexts from  
theoretical physics to the experimental
sciences and engineering applications. 
A priori, the phenomena of superoscillations 
can play a role in each of these contexts.
\sn
Our aim here has been to investigate the 
general properties of superoscillations. 
In particular, we found precise results about the extent to which 
frequency limited functions can superoscillate.    
Further, we gave a reliable characterization of the effect of
frequency limitation on the behavior of functions,
in terms of uncertainty relations. 
\sn
We formulated much of our discussion in
the concrete and intuitive language of information
theory but, of course, our results can easily be translated into all those
physical contexts where frequency limited functions occur. 
Here, we chose to always translate our results into the 
context of ultraviolet cut-off fields, 
where the ultraviolet cut-off is understood in the original
sense of a high frequency cut-off. We mentioned that
 superoscillations in field theory have been suggested,
 by Rosu and Reznik,  to
resolve the transplanckian frequencies paradox of black hole radiation.
In this context, our results showed that while 
generic superoscillations of arbitrarily high frequencies do exist,
they could be too instable under perturbations by interactions.  
This problem should be worth further pursuing.
\sn
We also obtained the general result that  
strictly bandlimited signals obey a
 lower bound $\Delta T_{min}$
on the standard deviation
of  their Nyquist rate samples and we generalized to time-varying
bandwidths.
In terms of ultraviolet cut-off quantum field theory these results
mean that the fields in ultraviolet cut-off field theories 
obey a formal minimum spatial
uncertainty $\Delta X_{min}$, where the minimum value of this
formal position uncertainty can be spatially varying.
In particular, we found that in ultraviolet cut-off 
quantum field theories 
the Nyquist rate for signals corresponds exactly 
to the in general spatially varying
density of local degrees of freedom.
\sn
So-far in our discussion we assumed 
this short-distance cut-off to arise
from the crude ultraviolet
cut-off obtained by cutting off high momenta.
However, interestingly, the same short-distance structure 
can also arise for example in theories with
effective Heisenberg uncertainty
relations which contain correction terms of the form:
\be
\Delta X \Delta  P ~\ge~\frac{\hbar}{2}\left(1+k
 (\Delta P)^2 +...\right),
\label{uc} \ee 
As is easy to verify, for a suitable small positive
constant $k$, Eq.\ref{uc} indeed yields a lower bound \be \Delta X_{min}
~=~ \hbar \sqrt{k}\ee 
which could be at a Planck- or at a string scale. 
This type of uncertainty relation
implies that the momentum stays unbounded. This means 
that the short-distance structure which we have here considered -
a formal finite minimum uncertainty in position -
is not tied to putting an upper bound to momentum.
Indeed, correction
terms to the uncertainty relations of the type of Eq.\ref{uc} have
appeared in various studies in the context of quantum gravity and
string theory. For reviews, see e.g. \cite{garay,witten}. For
recent discussion of potential physical
 origins of this type of uncertainty
relations see e.g. \cite{santiago,haro}. \sn
Quantum mechanical and quantum field theoretical
models which display such uncertainty relations have been 
investigated in detail. For example, the ultraviolet regularity
of loop graphs in such field theories has been shown. See
\cite{ak-jmp-ucr,ak-jmp,kmm}.
\sn
In work by Brout et al, \cite{brout}, it 
has been shown that this type
of short-distance cut-off without
energy-momentum cut-off, when built into quantum theory
could resolve the transplanckian energies
paradox of black hole radiation - without invoking superoscillations.
Since as we
now see, both, the approaches of Rosu and Reznik, \cite{rosu,reznik}, and of
Brout et al, \cite{brout},
 assume in fact the same short-distance structure it
should be very interesting to investigate their relationship. A 
recent reference in this context is  \cite{lubo}. \bn 
Finally,
we remark that it is not necessarily surprising that various
 different studies in quantum gravity
and in string theory have led to the same short-distance structure,
namely the short distance structure that arises for example from 
the uncertainty relation 
Eq.\ref{uc}.
In a certain sense it is not even surprising that the
same type of minimum uncertainty structure also appears in 
communication engineering: \sn This is because, as has been shown in
\cite{ak-erice}, in any theory, any real degree of freedom which is described by
an operator which is
 \it linear \rm
can only display very few types of
 short-distance structures.
The basic possibilities are continua,
 lattices and two basic types of unsharp
short distance structures, which have been named ``fuzzy-A"
and ``fuzzy-B". All others are mixtures of these. 
Technically, the unsharp real degrees of freedom
are those described by 
simple symmetric operators with nonzero (and, for the two types fuzzy-A and
fuzzy-B 
either equal or unequal) deficiency indices.
The ``time" degree of freedom of electronic signals is real, the
corresponding time operator $T$ therefore had to fall into this classification,
and among the few possibilities it happened to be of the type fuzzy-A.
\sn
But equally, we can consider for example in the matrix model of string theory
the coordinates of D0-branes. These are encoded in (the diagonal of) self-adjoint
matrices $X_i$. The quantization and the limit for the matrix size
$N\rightarrow \infty$
are difficult, but it is clear that the $X_i$ will eventually be operators
which are at least symmetric i.e. that their formal expectation values are real.
The short distance structure which these $X_i$ display will therefore
fall into this classification which we mentioned. Since there are
only these few basic possibilities continuous, discrete or ``fuzzy", they
 may well be found to be
of one of the fuzzy, types. 
\sn
In the present paper we have been
concerned with short-distance structures 
which are characterized by a formal finite minimum uncertainty.  
The classification given in \cite{ak-erice} shows that 
all such degrees of freedom are of the type fuzzy-A.
\sn 
We can therefore also view our present results on
superoscillations as clarifying
aspects of one of these very general classes of
short-distance structures of real degrees of freedom.
Our results on superoscillations translate in any theory
which contains unsharp degrees of freedom of this type.
\bn
\bf Acklowledgements: \rm The author is very
grateful to Haret Rosu for
bringing the issue of superoscillations to his
attention, and the author
is happy to thank John Klauder and Pierre Sikivie
 for their very valuable criticisms.

\end{document}